\documentclass[12pt]{article}
\usepackage{amssymb}
\begin{document}

\begin{titlepage}

                            \begin{center}
                            \vspace*{2cm}
        \large\bf Geometric variations of the Boltzmann entropy\\

                            \vfill

              \normalsize\sf    NIKOS \ \ KALOGEROPOULOS\\

                            \vspace{0.2cm}

 \normalsize\sf Department of Science\\
 BMCC - The City University of New York,\\
 199 Chambers St., \ New York, NY 10007, \ USA\\

                            \end{center}

                            \vfill

                     \centerline{\normalsize\bf Abstract}
\normalsize\rm\setlength{\baselineskip}{18pt} \noindent We perform
a calculation of the first and second order infinitesimal
variations, with respect to energy, of the Boltzmann entropy of
constant energy hypersurfaces of a  system with a finite number of
degrees of freedom. We comment on the stability interpretation
of the second variation in this framework.\\

                             \vfill


                             \vfill

\noindent\rule{8cm}{0.2mm}\\
\begin{tabular}{ll}
\small\rm E-mail: & \small\rm nkalogeropoulos@bmcc.cuny.edu\\
                  & \small\rm nkaloger@yahoo.com
\end{tabular}
\end{titlepage}


                            \newpage

\centerline{\sc  1. \ Introduction}

                            \vspace{3mm}

\normalsize\rm\setlength{\baselineskip}{18pt}

In theories describing systems with many degrees of freedom, field
theories being an example, it is of great interest to be able to
derive thermodynamic quantities from the microscopic dynamics. The
microscopic dynamics of a non-dissipative system is encoded in its
Hamiltonian description. In a statistical description of such a
Hamiltonian system, we trade the practically intractable
symplectic evolution on its phase space with a ``reasonable"
probability measure that describes some of the characteristics of
such an evolution. In systems in thermodynamic equilibrium  the
use of the microcanonical and the canonical distributions has
proved exceedingly successful during the last century and these
are the distributions with respect to
which we will be calculating the statistical averages in this paper.\\

The Boltzmann entropy has proved to be one of the most useful
thermodynamic potentials. The Boltzmann entropy is proportional to
the area of the total energy \ $E$ \ hyper-surfaces \
$\mathcal{M}_E$ \ on which the configuration space \ $\mathcal{N}$
\ of the Hamiltonian system can be foliated. Because of this very
direct geometric interpretation, we have chosen to analyze
variations of the Boltzmann entropy. Naturally, the different
response functions of such a thermodynamic system can be derived
in terms of appropriate variations of the Boltzmann entropy. The
most geometrically transparent, and at the same time physically
relevant, of such variations are the ones with respect
to the total energy \ $E$ \ of each hypersurface, to which we focus.\\

An outline of this paper is as follows: in Sections 2 and 3
respectively, we calculate the first and second order change of
the Boltzmann entropy under infinitesimal variations of the energy
\ $E$ \ (diffeomorphisms) of \ $\mathcal{M}_E$. \ We also provide
a physical interpretation of these results in terms of models
without kinetic terms (e.g. lattice models). In Section 4 we
present some conclusions and the comment on the relation of this
paper with similar works.\\

                          \vspace{5mm}


     \centerline{\sc 2. \ First order variation of the entropy}

                          \vspace{3mm}

Assume, following the ideas of Krylov [1] that \ $\mathcal{N}$, \
the configuration space of the system under study, is an
$n$-dimensional Riemannian manifold with metric \ $\tilde{g}$. \
The evolution of such a system is described by the geodesic flow
of \ $\tilde{g}$. \ Consider two diffeomorphic, infinitesimally
close, hyper-surfaces \ $\mathcal{M}_E$ \ and \
$\mathcal{M}_{E+\delta E}$  of \ $\mathcal{N}$ \ with
corresponding energies \ $E$ \ and \ $E+\delta E$ \ respectively.
Since the system is autonomous, its evolution can be described by
restricting our attention to \ $(\mathcal{M}_E, g)$. \ When the
system is coupled to a heat reservoir, because of the existence of
energy fluctuations \ $\delta E$ \ the system may find itself in
an ``adjacent" hypersurface \ $\mathcal{M}_{E+\delta E}$. \ If \
$\delta E \ll E$ \ and the system is not close to a phase
transition, then it is reasonable to expect that \ $\mathcal{M}_E$
\ and \ $\mathcal{M}_{E+\delta E}$ \ are diffeomorphic.\\

 The thermodynamic potential with the most
straightforward geometric interpretation is the Boltzmann entropy.
The Boltzmann entropy describes accurately the behavior of a
system, as long as the system evolution is ergodic [2] on its
configuration space \ $\mathcal{N}$. \ Since it is difficult to
prove, in practice, the ergodicity of a system of physical
significance starting from first principles, we proceed by
assuming  that the geodesic flow is ergodic [2] on both \
$\mathcal{M}_E$ \ and \ $\mathcal{M}_{E+\delta E}$ \ when they are
diffeomorphic to each other. Let \ $f: \mathcal{M}_E \rightarrow
\mathcal{M}_{E+\delta E}$ \ be such a diffeomorphism, and \ $X\in
T\mathcal{N}|_\mathcal{M}$ \ be the  vector field generating \
$f$, \ when restricted on \ $\mathcal{M}$. \ $X$ \ can be written
as \ $X=\frac{d}{dE}\bigg|_E$ \ and let \ $\mathfrak{L}_X$ \
denote the Lie derivative along \ $X$. \  We assume that the
metric on $\mathcal{M}_E$ is the induced metric \ $g$ \ from \
$\tilde{g}$ \ [3]. Let \ $\tilde{\nabla}$ \ and \ $\nabla$ \
indicate the Levi-Civita connections on \ $\mathcal{N}$ \ and \
$\mathcal{M}$ \ compatible with \ $\tilde{g}$ \ and \ $g$ \
respectively [3]. Then
\begin{equation}
  \tilde{g}(e_i, e_j) = g(e_i, e_j)
\end{equation}
All Latin indices take values \ $1$ \ through \ $m$ \ in the
sequel. Although \ $\mathcal{M}_E$ \ is obviously a codimension 1
sub-manifold of \ $\mathcal{N}$, \ during most of the calculation
we will be more general, namely we will assume that the dimension
of \ $\mathcal{M}_E$ \ is \ $m<n$ \ without necessarily \ $m=n-1$.\\

As is well known, the Boltzmann entropy of \ $\mathcal{M}_E$ \ is
given by
\begin{equation}
    S_E = k_B \ln vol \mathcal{M}_E
\end{equation}
where $k_B$ denotes the Boltzmann constant and
\begin{equation}
  vol \mathcal{M}_E = \int\limits_{\mathcal{M}_E} \sqrt{\det g} \ \ d^mx
\end{equation}
Then
\begin{equation}
    \delta S := S_{E+\delta E}-S_E = k_B \ln \frac{vol
    \mathcal M_{E+\delta E}}{vol \mathcal{M}_E}
\end{equation}
By keeping terms up to second order in \ $\delta E$ \ and upon
expanding the logarithm we find
\begin{equation}
   \delta S = \frac{k_B}{vol \mathcal{M}_E}
   \frac{d(vol\mathcal{M}_E)}{dE}\Bigg|_E \delta E +
   \frac{k_B}{2 \ vol \mathcal{M}_E}
   \left[ \frac{d^2 ( vol \mathcal{M}_E)}{dE^2}\Bigg|_E - \frac{1}{vol \mathcal{M}_E}
   \left( \frac{d (vol \mathcal{M}_E)}{dE}\Bigg|_E  \right)^2 \right] (\delta E)^2
\end{equation}
which can be re-expressed [3] as
\begin{equation}
\delta S = \frac{k_B}{vol \mathcal{M}_E}
   (\mathfrak{L}_X vol\mathcal{M}_E) \delta E +
   \frac{k_B}{2 \ vol \mathcal{M}_E}
   \left[ \mathfrak{L}_X\mathfrak{L}_X  vol \mathcal{M}_E - \frac{1}{vol \mathcal{M}_E}
   \left( \mathfrak{L}_X vol \mathcal{M}_E  \right)^2 \right] (\delta E)^2
\end{equation}
Let \ $\{e_i\}, \ i=1,\ldots, m$ \ be an orthonormal basis of \
$T\mathcal{M}$ \ with respect to \ $g$. \ Let
$\widetilde{\nabla}$, \ $\nabla$ \ denote the Levi-Civita
connections compatible with \ $\tilde{g}$ \ and \ $g$ \ on \
$\mathcal{N}$ and \ $\mathcal{M}$ \ respectively [3]. Then (1),(3)
give
\begin{equation}
 \mathfrak{L}_X vol \mathcal{M}_E = \int\limits_{\mathcal{M}_E}
 \mathfrak{L}_X \sqrt{\det \tilde{g}(e_i,e_j)} \ d^mx
\end{equation}
For any positive-definite symmetric matrix $A$, we use the
identity \ $\det A = \exp(Tr \ln A)$, where \ $Tr$ \ denotes the
trace of \ $A$, \ and we get
\begin{equation}
 \mathfrak{L}_X vol \mathcal{M}_E = \frac{1}{2} \int\limits_{\mathcal{M}_E}
   \sqrt{\det \tilde{g}(e_p, e_q)} \ \ \mathfrak{L}_X Tr \ln [\tilde{g}(e_i, e_j)] \ d^mx
\end{equation}
We observe that the Lie derivative \ $\mathfrak{L}_X$ \  and the
trace \ $Tr$ \ operations commute, and by using (1)
\begin{equation}
 \mathfrak{L}_X vol \mathcal{M}_E = \frac{1}{2} \int\limits_{\mathcal{M}_E}
   \sqrt{\det g(e_p, e_q)} \ \ Tr \mathfrak{L}_X  \ln [\tilde{g}(e_i, e_j)] \ d^mx
\end{equation}
which eventually gives
\begin{equation}
 \mathfrak{L}_X vol \mathcal{M}_E = \frac{1}{2} \int\limits_{\mathcal{M}_E}
   \sqrt{\det g(e_p, e_q)} \ \ [g(e_i, e_j)]^{-1}  \mathfrak{L}_X
   [\tilde{g}(e_i, e_j)] \ d^mx
\end{equation}
where repeated indices are summed. Since \ $g(e_i, e_j) \in
\mathbb{R}$ \ are functions on  \ $\mathcal{M}_E \subset
\mathcal{N} $ \ [3]
\begin{equation}
\mathfrak{L}_X[\tilde{g}(e_i, e_j)] = \widetilde{\nabla}_X
[\tilde{g}(e_i, e_j)]
\end{equation}
Then
\begin{equation}
 \widetilde{\nabla}_X [\tilde{g}(e_i, e_j)] = (\widetilde{\nabla}_X
 \tilde{g})(e_i,e_j) + \tilde{g}(\widetilde{\nabla}_X e_i, e_j) + \tilde{g}(e_i,
 \widetilde{\nabla}_X e_j)
\end{equation}
Since \ $\widetilde{\nabla}$ \ is the Levi-Civita connection
compatible with \ $\tilde{g}$, \ then [3] \ $(\widetilde{\nabla}_X
\tilde{g})(e_i,e_j)=0$ \ and  its torsion is zero, namely
\begin{equation}
\widetilde{\nabla}_X e_i - \widetilde{\nabla}_{e_i}X = [X, e_i]
\end{equation}
where \ $[X,e_i]$ \ denotes the Lie bracket between the vector
fields \ $X$ \ and \ $e_i$. \ The metric \ $\tilde{g}$ \ gives
rise to the orthogonal decomposition \ $X = X^\top + X^\bot$ \
into a tangential component \ $X^\top \in T\mathcal{M}_E$ \ and
into a normal component \ $X^\bot \in N\mathcal{M}_E$, \ where \
$N\mathcal{M}_E$ \ denotes the normal bundle of \ $\mathcal{M}_E$.
\ Then (12), (13) and linearity imply
\begin{equation}
  \widetilde{\nabla}_X [\tilde{g}(e_i, e_j)] =
   \tilde{g}(\widetilde{\nabla}_{e_i}X^\bot, e_j) +
   \tilde{g}([X^\bot, e_i], e_j) +
   \tilde{g}(\widetilde{\nabla}_{X^\top}e_i, e_j) + (i\leftrightarrow j)
\end{equation}
where  \ $(i\leftrightarrow j)$ \ indicates similar terms with \
$i$ \ and \ $j$ \ interchanged. The first term of the above sum is
the second fundamental tensor [3] \ $l_{X^\bot}(e_i, e_j)$. \ By
using (13) again, we get
\begin{equation}
  \widetilde{\nabla}_X [\tilde{g}(e_i, e_j)] =
   l_{X^\bot}(e_i, e_j) + \tilde{g}([X, e_i], e_j) +
   \tilde{g}(\widetilde{\nabla}_{e_i}X^\top , e_j) +
   (i\leftrightarrow j)
\end{equation}
Substituting (15) into (10) we find
\begin{equation}
\mathfrak{L}_X vol \mathcal{M}_E = \int\limits_{\mathcal{M}_E}
   \sqrt{\det g(e_p, e_q)} \ \ \left\{
   l_{X^\bot}(e_i, e_i) + \tilde{g}([X, e_i], e_i)
   + \tilde{g}(\widetilde{\nabla}_{e_i}X^\top, e_i)\right\} \ d^mx
\end{equation}
The determinant under the radical is equal to one, since \
$\{e_i\}$ \ is an orthonormal basis with respect to \ $g$. \ The
third term is, by definition, the divergence of \ $X^\top$. \ Then
by Green's theorem, we find
\begin{equation}
 \int\limits_{\mathcal{M}_E} \tilde{g}(\widetilde{\nabla}_{e_i}X^\top,
 e_i) = \int\limits_{\partial\mathcal{M}_E} g(X^\top, \nu) \ d\mu
\end{equation}
where \ $\nu \in T\partial\mathcal{M}_E$ \ represents the outward
unit normal on the boundary \ $\partial\mathcal{M}_E$ \ and \
$d\mu$ \ is the induced Riemannian measure on \
$\partial\mathcal{M}_E$. \ If \ $X$ \ is perpendicular to \
$\mathcal{M}_E$, \ i.e. if \ $X^\top$ \ is zero,  or if \
$\mathcal{M}_E$ \ is closed, then this term is trivially zero. Let
\ $c_i(s)$ \ denote the integral curve of \ $e_i$, \ i.e. \
$\frac{dc_i(0)}{ds}=e_i$ \ and let \ $c_i(s,E)$ \ be the one
parameter variation of \ $c_i(s)$ \ along $X$, i.e. \
$c_i(s,0)=c_i(s)$ \ and \ $\frac{\partial c_i(s,E)}{\partial
E}=X$. Then
\begin{equation}
  X e_i = \frac{\partial}{\partial E}\Bigg|_E
  \frac{\partial}{\partial s}\Bigg|_0 c_i(s,E) = e_i X
\end{equation}
so the second term of (16) is zero. If the above  conditions hold,
then (16) simplifies to
\begin{equation}
\mathfrak{L}_X vol \mathcal{M}_E = \int\limits_{\mathcal{M}_E}
   \sqrt{\det g(e_p, e_q)} \ \ l_{X^\bot}(e_i, e_i) \ d^mx
\end{equation}
Let \ $\beta = 1/k_BT$, \ as usual. If the system under study has
a constant extensive variable \ $\mathcal{V}$, \ e.g. volume, and
constant ``particle number" \ $\mathcal{N}$, \ then
\begin{equation}
   \frac{\partial S}{\partial E}\Bigg|_{\mathcal{V, N}} =
   \frac{1}{T}
\end{equation}
and using (6),(20) we obtain
\begin{equation}
    \beta \ = \ \frac{\int\limits_{\mathcal{M}_E}
    \sqrt{\det g(e_j, e_k)} \ \  l_{X^\bot}(e_i, e_i) \ d^mx}{\int\limits_{\mathcal{M}_E}
    \ \sqrt{\det g(e_p, e_q)} \ d^mx}
\end{equation}
which can be interpreted as the average of the mean curvature \
$Tr \ l_{X^\bot}$ \ with respect to the micro-canonical measure
\begin{equation}
    \rho = \frac{1}{vol \mathcal{M}_E}
\end{equation}
and (21) can  be rewritten as
\begin{equation}
      \beta \ = \ \langle \ Tr \ l_{X^\bot} \ \rangle
\end{equation}
where the average $\langle \ \rangle$ \ is taken over \
$\mathcal{M}_E$. \ Since \ $\beta >0$ \ then \ $\langle \ Tr \
l_{X^\bot} \ \rangle
>0$. \ An example where this condition is satisfied is when
\ $\mathcal{N} = \mathbb{R}^n$, \ with \ $\mathcal{M}$ \ being
diffeomorphic to the sphere \ $S^{n-1}$ \ and isometrically
embedded in \ $\mathbb{R}^n$. \ Then  \ $l_{X^\bot} > 0$ \
everywhere on \ $\mathcal{M}$, \ a fact which clearly guarantees
the positivity of (23). For this example, and because \
$\mathbb{R}^n$ \ is non-compact, we assume that either \ $X^\top =
0$, \ or all the functions on \ $\mathcal{M}_E$ \  have compact
support. Generally, however, we cannot exclude the possibility
 \ $\langle \ Tr \ l_{X^\bot}\ \rangle < 0$. \ In such case,
(23) loses its direct physical interpretation. One reason why \
$\langle \ Tr \ l_{X^\bot}\ \rangle < 0$, \  can be traced to the
lack of ergodicity of the geodesic flow on \ $\mathcal{M}_E$, \
which was assumed at the outset. Without such ergodic behavior,
the expression for the Boltzmann entropy (2) is reduced to just a
formal definition devoid of any physical meaning. This lack of
physical meaning is subsequently inherited to thermodynamic
relations like (20), where \ $T$ \ can no longer be identified
with the physical quantity ``temperature". An alternative
interpretation of (23), is as a constraint equation on the
possible choice of a metric \ $\tilde{g}$ \ describing the
evolution of the system on \ $\mathcal{N}$. \ In such an
interpretation, a metric \ $\tilde{g}$ \ resulting in \
$l_{X^\bot} < 0$ \ is not acceptable, on physical grounds.
Therefore, either  the metric \ $\tilde{g}$ \ used for the
description of the system should be modified, or in extreme cases,
one should take the more radical step of discarding the
model altogether.\\

If the model under consideration, however, describes a system
(lattice models are, frequently, such examples) for which there is
an upper bound in the possible energy, then negative temperatures
are theoretically allowed, in the definition of the partition
function of a canonical treatment. Such models should not,
obviously, contain any kinetic energy terms and this is reflected
on the choice of \ $\tilde{g}$ \ on \ $\mathcal{N}$ \ describing
their evolution. Systems being described by such models have been
experimentally observed [4] to be out of equilibrium, a fact which
puts in question, the suitability of the Boltzmann entropy in
describing them, especially when they are far from equilibrium.
For such models there is no constraint on the sign of (23), but to
acquire a
physical meaning, \ $T$ \ should be interpreted appropriately.\\

                          \vspace{5mm}


     \centerline{\sc 3. \ Second order variation of the entropy}

                          \vspace{3mm}

For  systems in which both positive and negative temperatures have
physical meaning [4], a limiting case occurs when the
microcanonical average (23) of the mean curvature is zero, \
$\langle \ l_{X^\bot}(e_i, e_i) \ \rangle \ =0$, \ which amounts
to \ $\beta = 0$  \ or \ $T$ \ being infinite. This requirement is
trivially fulfilled [4] when \ $\mathcal{M}_E$ \ is a totally
geodesic submanifold of \ $\mathcal{N}$, \ i.e. when \
$l_{X^\bot}(e_i, e_j) = 0$ \ or  when \ $\mathcal{M}_E$ \ is a
minimal submanifold of \ $\mathcal{N}$, \ i.e. when the mean
curvature \ $l_{X^\bot}(e_i,e_i)=0$. \ In such cases the Boltzmann
entropy remains invariant under the action of \ $X$ \ and (6)
gives
\begin{equation}
  \delta S = \frac{k_B}{2 \ vol \mathcal{M}_E}
   \left( \mathfrak{L}_X\mathfrak{L}_X  vol \mathcal{M}_E  \right) (\delta E)^2
\end{equation}
To perform this calculation we start by Lie-differentiating (16)
\begin{eqnarray}
 \mathfrak{L}_X\mathfrak{L}_X  vol \mathcal{M}_E & = &
   \int\limits_{\mathcal{M}_E}
    \left\{ \mathfrak{L}_X\sqrt{\det g(e_p, e_q)} \right\}
   \left\{ l_{X^\bot}(e_i, e_i) + \tilde{g}([X, e_i], e_i) +
   \tilde{g}(\widetilde{\nabla}_{e_i}X^\top, e_i) \right\} \ d^mx
   + \nonumber \\
   & &  \int\limits_{\mathcal{M}_E} \left\{ \mathfrak{L}_X
   \{l_{X^\bot}(e_i, e_i)\} + \mathfrak{L}_X\{ \tilde{g}([X, e_i], e_i)\}
   + \mathfrak{L}_X \{\tilde{g}(\widetilde{\nabla}_{e_i}X^\top, e_i) \} \right\}
    \sqrt{\det g} \ d^mx \nonumber
\end{eqnarray}
The first term of the right hand side is given by (16). Taking
into account (18), the definition of the second fundamental form,
and that \ $X = X^\top + X^\bot $, \ we find
\begin{eqnarray}
 \mathfrak{L}_X\mathfrak{L}_X  vol \mathcal{M}_E & = &
   \int\limits_{\mathcal{M}_E} \left\{ \tilde{g}(\widetilde{\nabla}_{e_i}X^\bot, e_i) +
   \tilde{g}(\widetilde{\nabla}_{e_i}X^\top, e_i) \right\}^2 \sqrt{\det g} \ d^mx \
   + \nonumber \\
   & &  \int\limits_{\mathcal{M}_E} \left\{
   \widetilde{\nabla}_X\{\tilde{g}(\widetilde{\nabla}_{e_i}X,
   e_i)\} + \widetilde{\nabla}_X \{\tilde{g}([X, e_i], e_i)\} \right\}
   \sqrt{\det g} \ d^mx \nonumber
\end{eqnarray}
which gives, after using the torsion-free condition (13) with (18)
and performing the covariant differentiations
\begin{equation}
 \mathfrak{L}_X\mathfrak{L}_X  vol \mathcal{M}_E  =
   \int\limits_{\mathcal{M}_E}
   \left\{ \{\tilde{g}(\widetilde{\nabla}_{e_i}X, e_i)\}^2 +
   \tilde{g}(\widetilde{\nabla}_X e_i, \widetilde{\nabla}_X e_i) +
   \tilde{g}(\widetilde{\nabla}_X\widetilde{\nabla}_X e_i, e_i) \right\} \
   \sqrt{\det g} \ d^mx
\end{equation}
By using  that \ $\widetilde{\nabla}$ \ is Levi-Civita with
respect to \ $\tilde{g}$, \ this can also be written as
\begin{equation}
   \mathfrak{L}_X\mathfrak{L}_X  vol \mathcal{M}_E  =
   \int\limits_{\mathcal{M}_E}
   \left\{ \{\tilde{g}(\widetilde{\nabla}_{e_i}X, e_i)\}^2 +
   \widetilde{\nabla}_X\{\tilde{g}(\widetilde{\nabla}_X e_i,
   e_i)\}\right\} \sqrt{\det g} \ d^mx
\end{equation}
If \ $\mathcal{M}_E$ \ is a minimal submanifold of \
$\mathcal{N}$, \ then (26) reduces to
\begin{equation}
 \mathfrak{L}_X\mathfrak{L}_X  vol \mathcal{M}_E  =
   \int\limits_{\mathcal{M}_E}
   \left\{ \{\tilde{g}(\widetilde{\nabla}_{e_i}X^\top, e_i)\}^2 +
   \widetilde{\nabla}_X\{\tilde{g}(\widetilde{\nabla}_X e_i,
   e_i)\}\right\} \sqrt{\det g} \ d^mx
\end{equation}
A further simplification occurs when \ $X$ \ is everywhere normal
to \ $\mathcal{M}_E$, \ i.e. when \ $X^\top = 0$. \ Then
\begin{equation}
 \mathfrak{L}_X\mathfrak{L}_X  vol \mathcal{M}_E  =
   \int\limits_{\mathcal{M}_E}
   \widetilde{\nabla}_X\{\tilde{g}(\widetilde{\nabla}_X e_i, e_i)\} \
   \sqrt{\det g} \ d^mx
\end{equation}
Substitution of (28) into (24) gives
\begin{equation}
\delta S = \frac{k_B \ (\delta E)^2}{2 \ vol \mathcal{M}_E}
   \int\limits_{\mathcal{M}_E}
   \widetilde{\nabla}_X\{\tilde{g}(\widetilde{\nabla}_X e_i, e_i)\} \
   \sqrt{\det g} \ d^mx
\end{equation}
which can be re-expressed, by using (22), as the microcanonical
mean
\begin{equation}
 \delta S = \frac{k_B (\delta E)^2}{2} \ \  \langle \
 \widetilde{\nabla}_X\{\tilde{g}(\widetilde{\nabla}_X e_i,  e_i)\} \
 \rangle
\end{equation}
or, equivalently, as
\begin{equation}
\frac{\partial^2 S}{\partial E^2} = \frac{k_B}{2} \ \  \langle \
 \tilde{g}(\widetilde{\nabla}_X e_i, \widetilde{\nabla}_X e_i) +
 \tilde{g}(\widetilde{\nabla}_X\widetilde{\nabla}_X e_i, e_i) \
 \rangle
\end{equation}
Since \ $\tilde{g}$ \ has positive signature, the first term of
the right hand side of (31) is positive or zero. For the same
reason, the operator in the second term is elliptic, thus it
eventually has positive eigenvalues. Then each side of (31) can
either be positive or negative, in general.  Because of (20), and
since the heat capacity \ $C_{\mathcal{V}}$ \ under the constant
extensive variable \ $\mathcal{V}$ \  is
\begin{equation}
 C_{\mathcal{V}} = \frac{\partial E}{\partial
 T}\Bigg|_{\mathcal{V,N}}
\end{equation}
 (30) can be re-written as
\begin{equation}
  \frac{k_B\beta^2}{C_{\mathcal{V}}} = - \frac{1}{2} \ \langle \
 \widetilde{\nabla}_X\{\tilde{g}(\widetilde{\nabla}_X e_i,  e_i)\} \
 \rangle
\end{equation}
During the second order variation \ $\beta =0$ \ which, according
to (33), implies that either \ $C_{\mathcal{V}} = 0$ \ or \ $
\langle \
 \widetilde{\nabla}_X\{\tilde{g}(\widetilde{\nabla}_X e_i,  e_i)\} \
 \rangle = 0 $. \  In order to avoid extending the
expansion (6) to cubic and higher order terms in \ $\delta E$, \
we consider only the former option. The result of (33) has a
physical interpretation as long as
\begin{equation}
 \lim_{T\rightarrow +\infty} \ \frac{\beta^2}{C_{\mathcal{V}}} =
 \widetilde{C}
\end{equation}
is finite. If \ $ \langle \
\widetilde{\nabla}_X\{\tilde{g}(\widetilde{\nabla}_X e_i,  e_i)\}
\ \rangle < 0 $ \ then \ $C_\mathcal{V} > 0$ \ which is the
standard stability criterion. On the other hand, if \ $ \langle \
\widetilde{\nabla}_X\{\tilde{g}(\widetilde{\nabla}_X e_i,  e_i)\}
\ \rangle > 0$, \ then \ $C_\mathcal{V} < 0$ \  which indicates
that the system is unstable. We can, therefore, interpret \ $
\langle \ \widetilde{\nabla}_X\{\tilde{g}(\widetilde{\nabla}_X
e_i, e_i)\} \ \rangle $ \ as a quantitative measure of the
instability of a system with an upper bound on its energy. \\

                           \vspace{5mm}


\centerline{\sc 4. \ Discussion and conclusions}

                           \vspace{3mm}

Some of the above results are standard in the theory of minimal
submanifolds [5],[6],[7] and the theory of harmonic maps [3]. In
the second variation, we deviated considerably from the
established practice [6],[7] which results in an inner product of
\ $X$ \ with an elliptic operator (Jacobi operator) expressed in
terms of the Laplacian and of the Riemann tensor of the normal
bundle \ $N\mathcal{M}_E$ \ acting on \ $X$. \ We did so because
we did not need the aforementioned geometric result in order to
obtain a physical interpretation of the second order variation of
the entropy. Evidently our result can be recast in the form
provided by [5],[6],[7] upon integration by parts and by using
(23) with \ $\beta = 0$.\\

It may also be worth noticing the similarity of the present
results to the ones of [8],[9]. In these papers the author relies
mostly on measure-theoretical arguments to draw his conclusions.
The use of a Euclidean metric on the phase space is very minimal
[8] to none [9]. Clearly, measure-theoretical arguments
[2],[9],[10],[11] are applicable to a much wider variety of
systems than the mechanical Hamiltonian ones that we use here.
Dissipative systems [10],[11] are an important class of systems
that the Riemannian approach, as used in the present paper, cannot
describe. On the other hand, the Riemannian approach may shed some
light into aspects of Hamiltonian systems as, for instance, the
relation between Gaussian curvature and dynamical temperature [8]
which may not be so clear, or accessible, if one uses purely
ergodic arguments. Such a relation has been pointed out by the
author of [8], who curious as he was about it, made no attempt to
trace its origins. Whether such a relation actually exists and can
be elucidated by using  Riemannian methods can be a
topic of future research.\\


                               \vspace{5mm}

\centerline{\normalsize\sc References}

                                  \vspace{3mm}

\noindent\rm [1] N.N. Krylov, \ \emph{Works on the
Foundations of Statistical Mechanics,} \\
       \hspace*{4.5mm} Princeton Univ. Press (1979).\\
\noindent\rm [2] A. Katok, B. Hasselblatt, \ \emph{Introduction
                  to the Modern Theory of Dymanical systems,}\\
                  \hspace*{6.5mm}  Camb. Univ. Press (1997).  \\
\noindent\rm [3] J. Jost, \ \emph{Riemannian Geometry and
Geometric Analysis,} \ 2nd Ed., \ Springer (1998).  \\
\noindent\rm [4] E.M. Purcell, R.V. Pound, \ \emph{Phys. Rev.} \ {\bf 81} \ 279 \ (1951). \\
\noindent\rm [5] A. Duschek, \ \emph{Math. Z.} \ {\bf 40} \ 279 \ (1936).\\
\noindent\rm [6] J. Simmons, \ \emph{Ann. Math.} \ {\bf 88} \ 62 \ (1968).\\
\noindent\rm [7] H.B. Lawson Jr., \ \emph{Lectures on Minimal
                  Submanifolds,} \ Vol.1 \ Publish or Perish (1980).\\
\noindent\rm [8] H.H. Rugh, \ \emph{Phys. Rev. Lett.} \ {\bf 78} \ 772 \ (1997).\\
\noindent\rm [9] H.H. Rugh, \ \sf{arXiv:chao-dyn/9703013}\\
\hspace*{6.5mm} \rm  H.H. Rugh, \ \emph{Phys. Rev. E} \ {\bf 64} \
055101 \ (2001).\\
\noindent\rm [10] D. Ruelle, \ \emph{J. Stat. Phys.} \ {\bf 95} \  393 \ (1999).\\
\hspace*{6.5mm}   G. Gallavotti, D. Ruelle, \  \sf{arXiv:chao-dyn/9612002}\\
\hspace*{6.5mm} \rm  D. Ruelle, \ \emph{AMS Bull.} \ {\bf 41} \ 275 \ (2004).\\
\noindent\rm [11] L.S. Young, \emph{J. Stat. Phys.} \ {\bf 108} \ 733 \ (2002). \\

                         \vfill

\end{document}